# Electromagnetic polarization controlled perfect switching effect with high refractive index dimers. the beam-splitter configuration

Ángela I. Barreda[1], Hassan Saleh[2,3], Amelie Litman[2], Francisco González[1], Jean-Michel Geffrin[2], Fernando Moreno[1*]

[1]*Group of Optics. Department of Applied Physics. University of Cantabria, 39005 Spain.*

[2]*Aix-Marseille Université, CNRS, Centrale Marseille, Institut Fresnel UMR 7249, 13013 Marseille, France.*

[3]*Centre Commun de Ressources en Microondes CCRM, 5 rue Enrico Fermi, 13453 Marseille, France.*

High Refractive Index (HRI) dielectric particles smaller than the wavelength, isolated or forming a designed ensemble are ideal candidates as new multifunctional elements for building optical devices. Their directionality effects are traditionally analyzed through forward and backward measurements, even if these directions are not suitable for practical purposes. Here we present unambiguous experimental evidence in the microwave range that, for a dimer of HRI spherical particles, a "perfect" switching effect ("perfect" means "off" = null intensity) is observed out of those directions as a consequence of the mutual particle electric/magnetic interaction. The "binary" state depends on the excitation polarization ("polarization switching"). Its analysis is performed through the linear polarization degree of scattered radiation at a detection direction perpendicular to the incident direction: the *beam-splitter* configuration. The scaling property of Maxwell equations allows generalizing our results to other frequency range and dimension scales, for instance, the visible and the nanometric scale.

One of the most active research branches relative to the interaction of electromagnetic radiation with matter is that of its scattering by particles smaller than the wavelength of the incident radiation. During many years, it has undergone vigorous investigation leading to applications in areas as diverse as health, material analysis, communications, etc[1]. For the case of metallic materials, when an impinging radiation illuminates a small metallic particle, the electronic plasma oscillates with the same frequency as the incident radiation. When resonant conditions are achieved, Localized Surface Plasmons (LSPs) are generated. These coherent oscillations of the metallic free electrons depend on the optical properties of the particle, its surrounding medium and also on the particle size and shape and the wavelength of the incident radiation[2]. When LSPs are generated, the energy of the impinging radiation is transferred to free electrons, which oscillate at maximum amplitude, and enhancements of the electric field are observed in the particle surroundings as well as strong electromagnetic energy localization. In spite of the strong response of metallic materials, like gold and silver in infrared and visible spectral ranges, their inherent ohmic losses make them less attractive for some particular applications, among which those concerning optical communications[3].

High Refractive Index (HRI) dielectric particles with low absorption have been proposed as an interesting alternative to overcome these losses issues[4-11]. Some of their most important advantages are related to the fact that light can travel through them without being absorbed and their compatibility with well-known technologies as they can be made with classical semi-conductors like silicon. In addition, they can be designed to control the direction of the scattered radiation. Under some specific conditions, known as Kerker conditions[12-18], scattered radiation obtained from a single HRI spherical particle can be concentrated either in the back or forward scattering regions. For the latter, it is even possible to produce a null scattering effect in the exact backward direction. These effects, which in the literature are called magnetodielectric, are a consequence of the coherent effects between the excited electric and magnetic dipolar modes. In general, whispering gallery modes are responsible for resonances involved in these phenomena, which can be either electric or magnetic, even though the particle is non-magnetic at all[5]. Their coherent contribution produces interference effects leading to "peculiar" directionality phenomena which can result in various applications for optical communication purposes. As such, very recently, it has been shown that small dielectric HRI spheres can be proposed as new multifunctional elements for building optical devices[19-21]. This can be achieved by taking advantage of the coherent effects not only between dipolar contributions but also between these and modes of higher order, giving rise to *anomalous* scattering effects[22]. These open an extra way to control the directionality of the scattered light.

By pursuing this idea, here we introduce the possibility of using a HRI dimer as an elementary unit for building binary switching devices. Although this possibility has also been recently proposed[23], back and forward directions are not really suitable for practical purposes for building operational optical circuits. In the forward direction, the scattered (wanted) signal is mixed with the incident beam while having a much lower intensity (for example, in our experiments, in the forward direction, the intensity of the incident field is more than thousand times higher than the scattered one). In the backward radiation, the scattered field is not easy to isolate without using some sort of beam-splitter with the corresponding complication in designing the scattering arrangement and as it can be shown, it is not possible to get perfect switching even if the two spheres do interact. Thus other directions, different from back and forward, seem more appealing for practical optical circuits design. In this work, we show that the scattered intensity at 90° from the incident direction can be null or maximum by playing with the polarization of a single frequency excitation and with a dimer whose components are close enough to interact in a controlled way. This interesting behavior converts the dimer element in a two-output (beam splitter) switching unit whose "binary" state depends only on the polarization of the exciting radiation. It is also remarkable that this switching effect is produced by the spectral evolution of one of the natural resonances of the isolated particle to an asymmetric shape resonance (Fano-like) as the particle interaction increases and the consequent implications in the practical use of this kind of spectral asymmetric shapes[24-27]. This means that the control of the electromagnetic interaction in a dimer element permits not only the proposal of a new binary (switching) unit but a control of the spectral properties of the scattered radiation.

The analysis of this "polarization switching" effect is performed here through the determination of the polarimetric parameter *$P_L(90º)$*, i.e. the linear polarization degree of the scattered radiation at a direction of detection which is perpendicular to the incident direction. Moreover, in order to finely understand the underlying physics associated to this polarization switching effect, we analytically study the different contributors by means of a Green's function formalism and a dipolar approximation. As this is an efficient tool to describe the electromagnetic interaction between particles, it enables to highlight the key parameters which are responsible for the expected behavior. We also demonstrate that this effect is perfectly tunable by simply changing the particle sizes.

Finally, it is important to remark that although the experiments contained in this research are made in the microwave range, the scale property of Maxwell equations permit to generalize their results to other dimensions and spectral ranges, including those of the nanoscale and the visible region. Also, they pave the way to design and build new optical elements to perform logical operations with light.

**Results**

The interaction effect between the two components of a homogeneous dimer has been analyzed for two identical spherical particles of radius $R_1=R_2=R$ and made of a high refractive index (HRI) material ($Re(n) > 3$). The considered particle is smaller than the illuminating radiation wavelength $\lambda$ (we define the size parameter $q$ as $2\pi R/\lambda$). The dimer is oriented along the Y-axis. This structure is excited by a linearly polarized plane wave propagating along the Z-axis and polarized either parallel to the dimer principal axis (longitudinal configuration) or perpendicular to it (transverse configuration), as it is shown in Fig. 1. The *gap separation distance* between the two external boundaries of the spheres is denoted by *d*. In order to express the gap distance in a dimensionless way, we define a new relative parameter, $d_0$, as the distance between the two particles, *d*, divided by the particle radius. This parameter is an indicator of the strength of the electromagnetic interaction between the two components of the dimer. Two gap distances are considered, $d_0=1/3$ (*small gap*→strong interaction) and 2 (*large gap*→weak interaction).

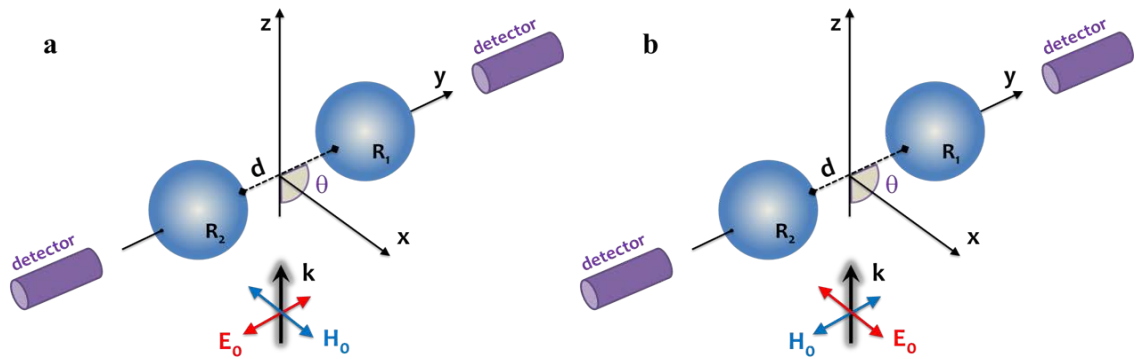

**Figure 1 | Scattering configuration**. A homogeneous sphere dimer of radius R is illuminated by a monochromatic plane wave propagating along the -Z-axis and linearly polarized either parallel to the dimer connecting axis ((**a**) longitudinal configuration) or perpendicular to it ((**b**) transverse configuration). The gap distance is denoted by *d* and the interaction parameter by $d_0=d/R$. $\theta$ corresponds to the scattering angle.

**Linear polarization degree for a High Refractive Index dimer.** The linear polarization degree, $P_L(\theta)$, can be defined[28-31], through the scattered intensities at a scattering angle $\theta$, when the exciting radiation is linearly polarized, oriented perpendicular and parallel to the scattering plane, as :

$$P_L(\theta) = \frac{I_S(\theta) - I_P(\theta)}{I_S(\theta) + I_P(\theta)} \qquad (1)$$

$I_S(\theta)$ (resp. $I_P(\theta)$) is the scattered intensity at a scattering angle $\theta$, when the exciting radiation is linearly polarized and perpendicular (resp. parallel) to the scattering plane (ZY in Fig. 1), which corresponds here to the transverse configuration (resp. longitudinal configuration). As it was demonstrated in previous studies, this polarimetric parameter $P_L$ is an efficient alternative to the more conventional extinction efficiency determination in light scattering experiments for particle sensing and sizing[13,32-34]. Furthermore, it can provide information about either the electric or magnetic nature of resonances[13].

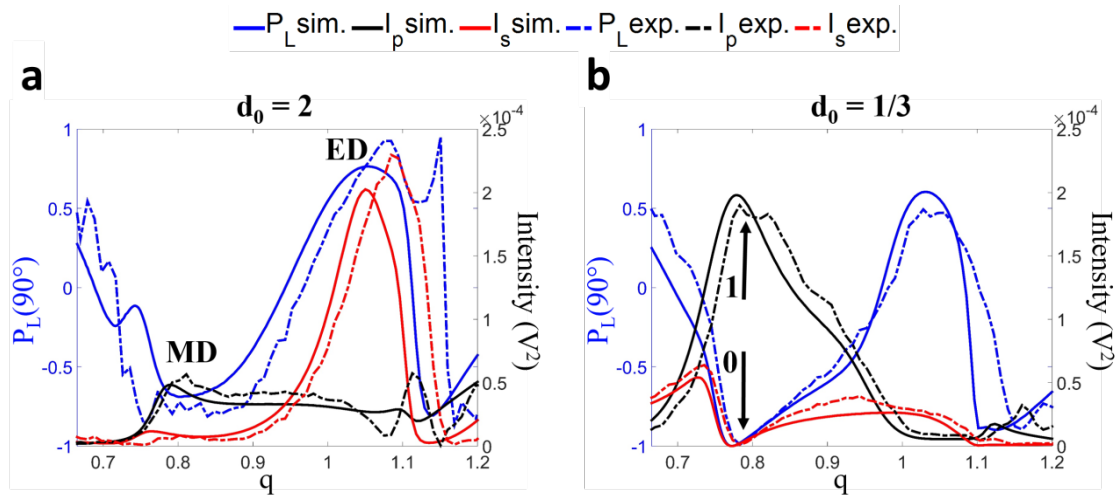

**Figure 2 | Polarimetric spectral measurements.** Linear polarization degree at right angle scattering configuration $P_L(90°)$ and scattered intensities from simulations and experiments at 90º for a sphere dimer as a function of the size parameter $q$. The exciting radiation is linearly polarized and the electric field is either perpendicular ($I_S$) or parallel ($I_P$) to the scattering plane (see Fig. 1). **(a)** Simulation and

measurement for $d_0$=2. **(b)** Simulation and measurement for $d_0$=1/3. Simulated $P_L$ (solid blue line) and measured $P_L$ (dashed blue line). Simulated $I_S$ (solid red line) and measured $I_S$ (dashed red line). Simulated $I_P$ (solid black line) and measured $I_P$ (dashed black line). The low and high states are indicated are indicated with the 0 and 1 arrows respectively.

In Fig. 2, we plot the spectral behavior of $I_S$, $I_P$ and the linear polarization degree $P_L$ at right angle scattering configuration, $\theta = 90°$, for the small and large gap cases in the dipolar spectral region where only the electric and magnetic dipolar contributions exist. It is possible to distinguish the dipolar electric (ED) and magnetic (MD) resonances, labeled in Fig. 2 as ED and MD respectively. Comparing the small and large gap cases, one of the most remarkable differences is the values of $P_L(90°)$ around $q$=0.8. For the small gap case, $P_L(90°)$ reaches -1, whilst for the large one, it clearly deviates from this value. According to Eq. (1), it is clear that $P_L(90°)$=-1 corresponds to null values of $I_S(\theta)$ and, as shown in [28], in the dipolar region, negative values of $P_L(90°)$ are linked to magnetic contributions. Around $q$=0.8, as the distance between the particles decreases, the intensity $I_P$ is increasing while $I_s$ is decreasing. In particular, $I_S$ reaches null values for $d_0$=1/3 at $q$=0.773 whilst $I_P$ reaches simultaneously its highest value. The reason of this behavior is twofold, as analyzed in the Supporting Information. First, the coupling between the two particles generates a supplementary induced magnetic dipole orientated along the propagation direction, when the transverse configuration is considered. Second, even if the induced electric and magnetic dipoles are less than $\lambda/2$ apart, they both generate electric fields which destructively interfere at $q$=0.773 for 90º scattering angle. Indeed, at that frequency, the interference of the electric field created by the electric and magnetic dipoles in both particles is destructive as their phase difference is $\pi/2$. Furthermore, the amplitude of the electric field created by the electric and magnetic induced dipoles in the first particle is the same as that created by the induced dipoles in the second particle. Thus, at that particular frequency, the dimer as a whole works as a pure magnetic unit ($P_L(90º)=-1$) leading to a "perfect" switching.

The fact that, at the same spectral position, denoted in the following as the "switching" frequency, $I_P$ and $I_S$ take so different values has important consequences for switching purposes. When both components of the dimer are close enough to interact between them, the ensemble behaves as a polarization switching element. This switching behavior is observed for different scattering angles. However, we operate at 90º for two main reasons. First, the biggest difference between the parallel and perpendicular to the scattering plane intensities is obtained for the right angle scattering configuration, which means that a clear difference between two states of switching device can be easily detected. Second, detecting at 90º is a good way to avoid any parasitic effect due to the incident radiation (considering a reasonably focused incident beam) making this dimer geometry and the incoming wave probably the best configuration for this switch to be of great practicality. Another remarkable feature is that we can obtain information about the charge distributions in the particles, observing the existence of a pure magnetic dipole at the switching frequency through the polarimetric parameter $P_L(90º)$. In general, out of the forward and backward directions, we have an additional feature due to the symmetry of the electromagnetic problem: two identical outputs can be handled at the same time with the same scattering unit and for one incidence. This feature is similar to that of a "beam splitter". Although this switching effect can be also observed for an isolated sphere, the dimer presents two main advantages. First, the difference between the high and low (or "on"/ "off") states is higher than with a single sphere. Second, the *low state is zero* (without noise), which makes it easier to detect and the *high state reaches higher values* than for an isolated sphere.

In Fig. 3, the scattering diagrams for both polarizations and for the two gaps are shown in the scattering plane ZY in the spectral range of the switching frequency. With the small gap, a large difference is observed between the scattered intensity values at 90º when comparing the two considered polarizations. However, for the large gap, the scattered intensities in the two polarization cases are both very small, which means that this switching effect is not observable at this wavelength when the particles do not interact.

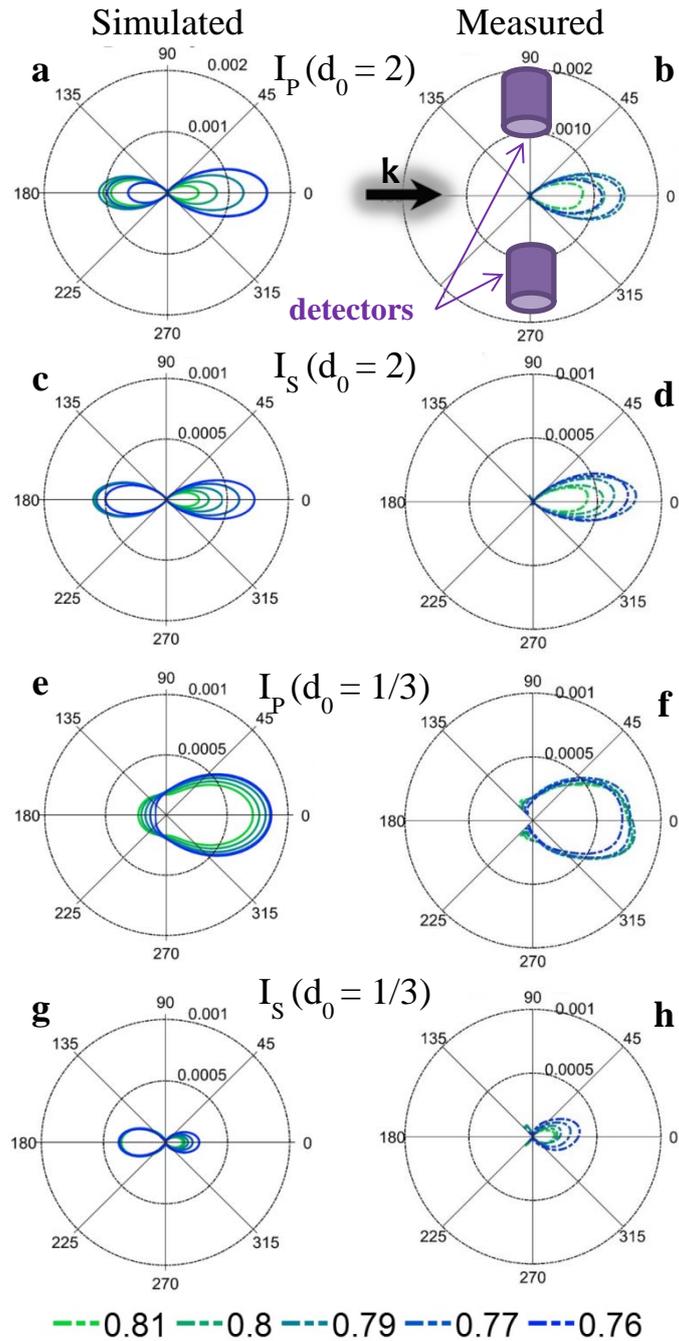

**Figure 3 | Scattered intensities angular variations.** Polar plot of the scattering intensities in the scattering plane around the "switching" frequency $q$=0.773. **(a,b,e,f):** Longitudinal ($I_P$) configuration. **(c,d,g,h):** Transverse ($I_S$) configuration. **(a-d):** Large gap, $d_0$=2. **(e-h):** Small gap, $d_0$=1/3. **(a,c,e,g):** Simulated intensities. **(b,d,f,h):** Measured intensities. It can be noticed that the measured angular range does not include backward scattering due to experimental restrictions (losing 100º backward). The black arrow indicates the direction of the incident radiation. The purple cylinders represent the detectors positions for the *beam-splitter configuration*.

To have a closer view at the field interaction in the vicinity of the dimer, the near field maps of the electric field intensity ($|\vec{E}|^2$) are plotted in Fig. 4 for the small gap case. A major difference in the intensity values is observed between the two polarization cases. For the longitudinal ($I_P$) polarization (Fig. 4ac), a hot spot can be observed in the gap. However, for the transverse ($I_S$) polarization (Fig. 4bd), the intensity values are negligible compared to the previous ones. The field evolution in near-field is thus in perfect agreement with the one observed in far-field and shown in Fig. 2.

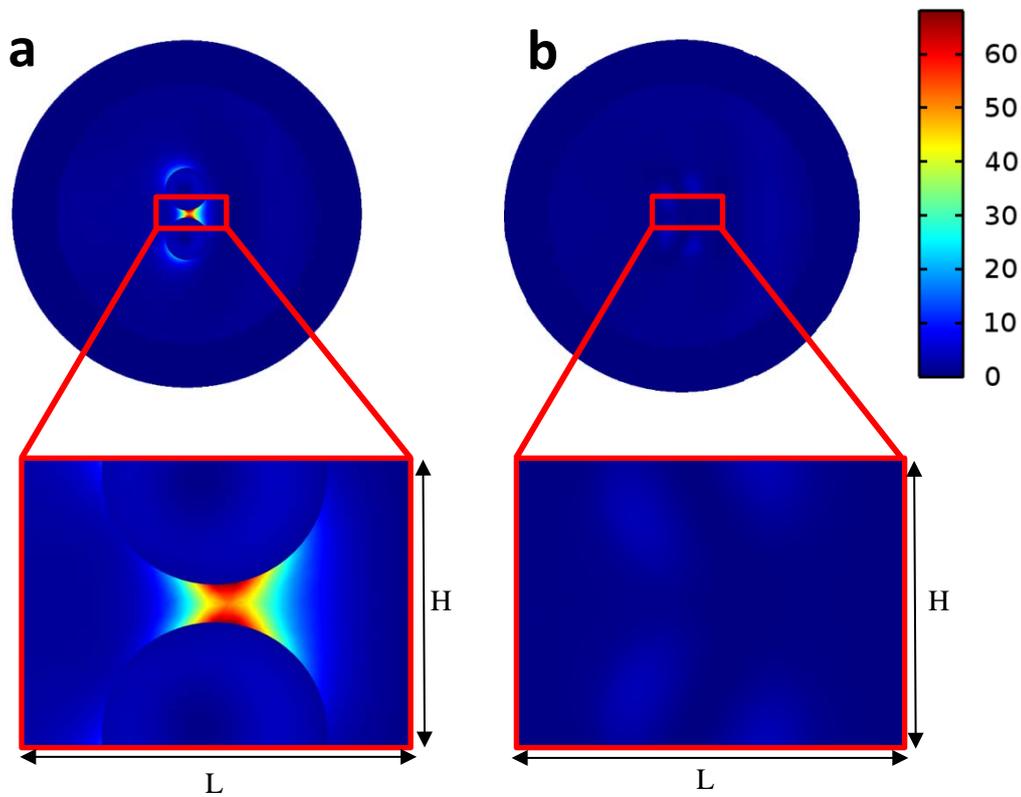

**Figure 4 | Near field intensity distributions.** Near field numerical maps of the electric field intensity ($|\vec{E}|^2 \frac{V^2}{m^2}$ in linear scale) in the scattering plane (ZY in Fig. 1) for the **(a)** longitudinal and **(b)** transverse configurations at the "switching" frequency $q=0.773$. Only the small gap case, $d_0=1/3$, is represented. The external annular domain corresponds to the PML layer. A closer view of the electric field intensity inside the gap is provided in the zoomed region corresponding to the rectangle drawn in (a) (resp. (b)), (L = 28mm, H = 21 mm).

In practical conditions, a switch must be built in order to be performing even within the most challenging conditions. With this in mind, the experiments described in Fig. 5 have been made with a rather focused incoming wave (as a laser beam can be), in order to fully assess the potential of dimers as optical switches. Those experiments are presented here without any kind of processing and are definitely raw measurements, as it would be with an actual device. Indeed, Fig. 5c and Fig. 5d are showing the intensity acquired as they come out from the receiver when measured with or without the dimer (the signal is even not referenced to the signal delivered to the source antenna, and no frequency variation is compensated). Only Fig. 5b (similar to Fig. 2b but without the $P_L$ curve) is obtained through the subtraction of these two raw complex measurements. Notice that the small differences between the measured fields in the three experiments with the P polarization observed in Fig. 5d are mainly due to the antennas imperfection, which are visible in Fig. 5c. Fig. 5d definitely provides a scientific evidence that a simple measurement of the intensity seen at ±90° when changing the polarization of the incoming wave can undeniably be used as the "on" and "off" states of a switching device. Furthermore, even if those measurements are made in the GHz range, this last proof of concept made with raw intensities really paves the way to exploit such a device in optics.

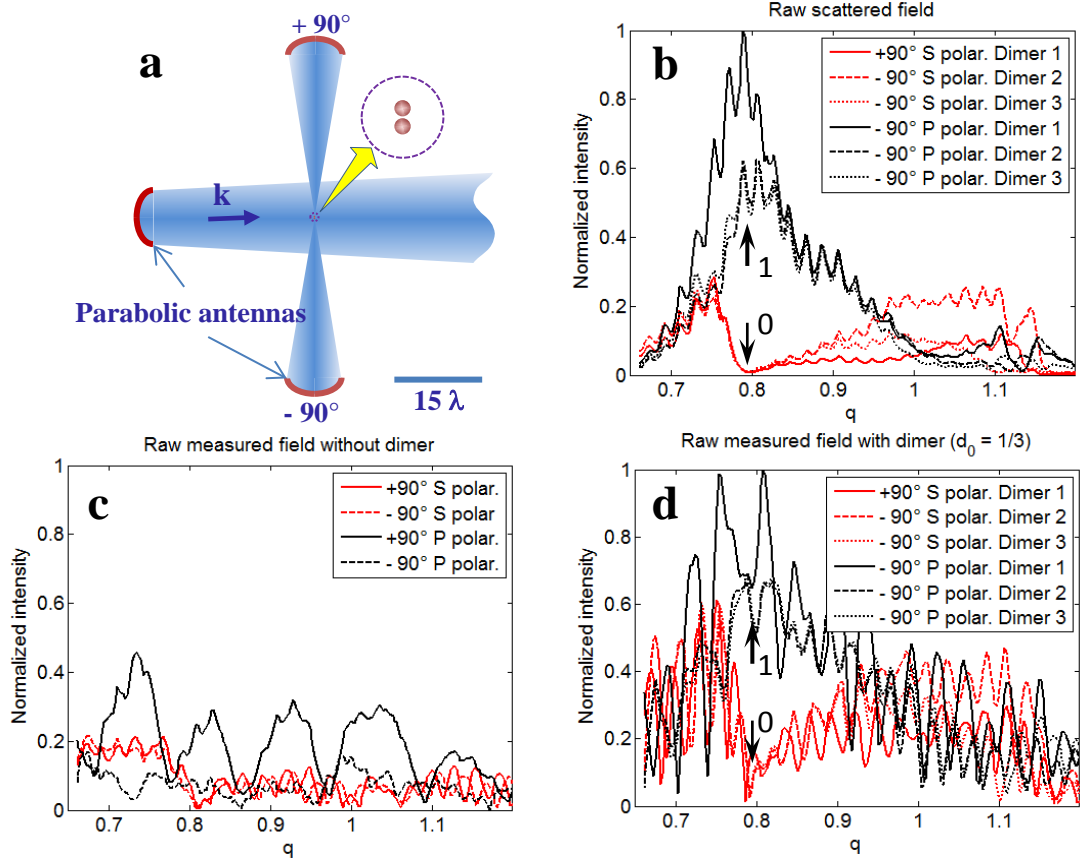

**Figure 5 | Direct measurements of the switch states.** (**a**) Experimental setup with directive antennas. (**b**) Raw measured scattered fields (without any post-processing) for receivers located at ±90° and for 3 different dimers (3 different associations of four spheres made of the same material and with identical diameters). (**c**) Measurements without any dimer to indicate the level of the residual signal due to the direct radiation between the antennas. (**d**) Detection of the useful field showing that the "on" and "off" states of the switching device can be directly measured without any reference nor processing.

**Tuning the switching frequency.** One of the most important characteristics of a device is the reproducibility of its performance in different spectral ranges. In particular, for our switching device, the "switching" frequency is perfectly tunable to different wavelengths by only changing the size of the particles. In this section, we consider two different radii, $R_1=R_2=R=6$ mm and $R_1=R_2=R=12$ mm. The gaps $d$ are selected such that the gap/particle size ratios $d_0$ take the same values as the ones studied in the previous section. In Fig. 6 we represent $I_S$, $I_P$ and $P_L(90°)$ for the new particle sizes, as well as for the previous one for the sake of comparison. Taking into account that the switching effect is observed in the dipolar region, the results shown are carried out using the analytical solution by means of the dipolar approximation without loss of generality. For the small gap, the particles are close enough to interact between them. For the small gap $d_0=1/3$, the switching effect is observed at shorter (resp. longer) wavelengths for $R=6$ mm (resp. $R=12$ mm) than for $R=9$ mm. This behavior is expected as the spectral position of the resonance depends on the particle size. As the particle size increases/decreases, the resonances are red/blue shifted[35]. In fact, when the electric permittivity value is higher than 12, it can be shown that there is a single set of *universal values* ($d_0$, $q$) which will ensure the null value for the scattered intensity at 90º for the transverse configuration (see Supplementary Information). The radius value $R$ has thus a direct connection with the "switching" frequency range and can act as the tuning parameter to control in order to achieve the required "switching" frequency. Also, as $R$ increases, the switching high state level increases, leading to a more efficient logical device.

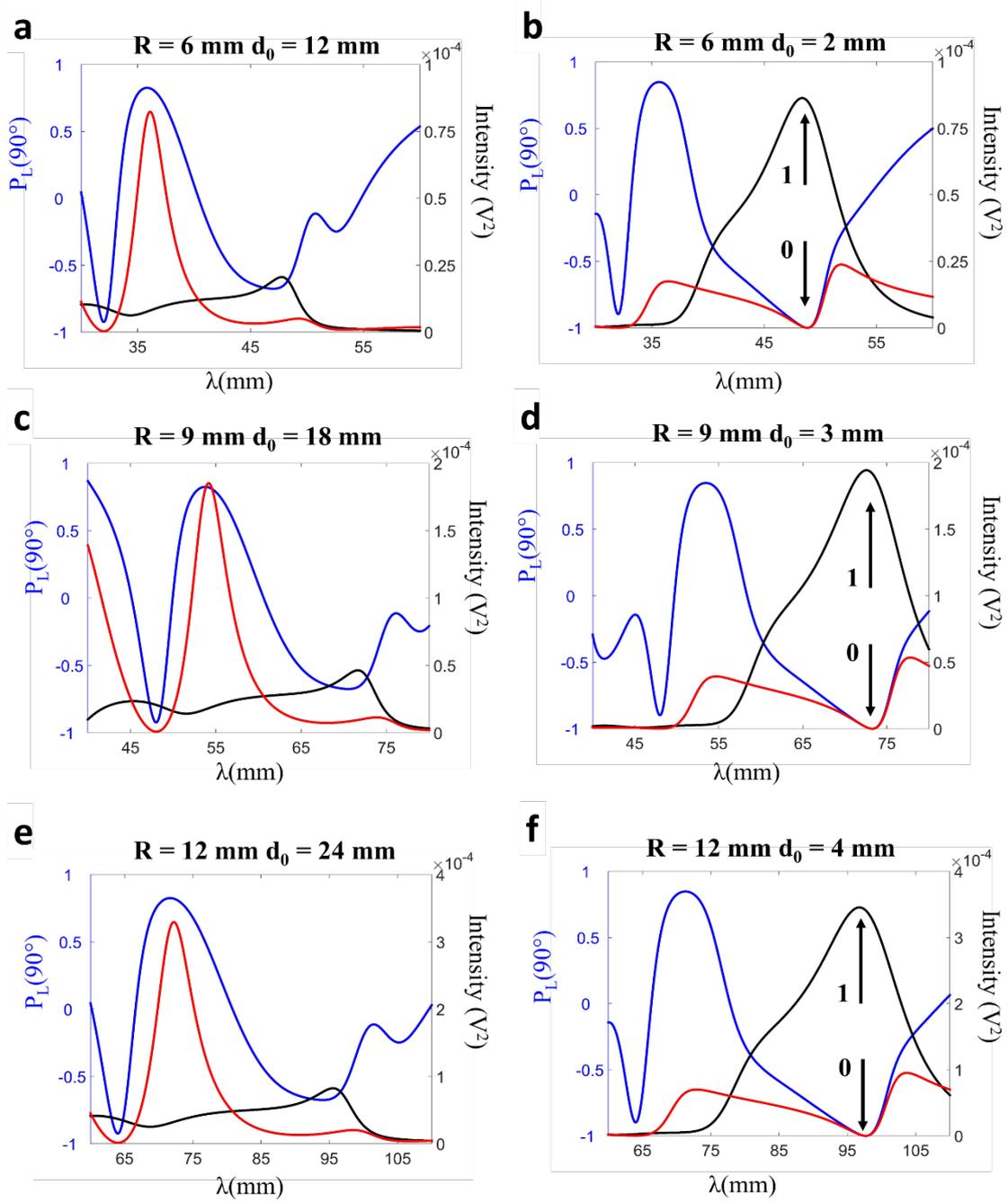

**Figure 6 | Spectral evolution of the switching frequency.** Linear polarization degree, $P_L(90°)$ and scattered intensities $I_S$ and $I_P$ measured at $90°$ in the scattering plane for a sphere dimer. Three types of dimer with identical spheres are considered: **(a,b)** $R_1=R_2=R=$6 mm, **(c,d)** $R_1=R_2=R=$9 mm and **(e,f)** $R_1=R_2=R=$12 mm. Two gaps are analyzed: **(a,c,e)** $d_0=$2 and **(b,d,f)** $d_0=$1/3. The black arrows represent the "on" (1) and "off" (0) states of the switching device. The simulations are performed with the dipolar model.

**Discussion**

We have introduced theoretically and demonstrated experimentally a practical arrangement for using a homogeneous dimer of high refractive index dielectric spheres as a perfect binary polarization controlled switching device. This has been done by acquiring the scattered intensities at a right angle scattering configuration. This can be considered equivalent to a 50/50 beam-splitter configuration which produces two identical beams from an only incident one. It has also been shown that this arrangement leads to the highest offset between the two polarization switching states (in our case, 0 state corresponds to null intensity since the dimer behaves as a perfect magnetic scatterer), while being out of the more inconvenient but classical forward and backward directions which are fully inoperative. We have also shown that its "binary" state depends only on the polarization of the exciting radiation. The possibility of tuning the "switching" frequency at which this phenomenon is the most significant has also been analyzed. A wide spectral tuning range can be obtained by modifying the particle size of the dimer components.

Moreover, under the dipolar approximation, an analysis based on the linear polarization degree and on the Green's function formalism, has enabled us to understand the physics of this electromagnetic problem and to optimize its behavior. First, the presence of a supplementary magnetic dipole generated by the coupling effect between the two particles, and second, the interference effect of the electric and magnetic dipoles induced by the incident beam in both particles of the dimer. We have also reported for the first time a pure magnetic dipole behavior at the specific "switching" frequency, i.e. at the right angle configuration the dimer as a whole works as a pure magnetic unit. This being the physical reason leading to a perfect switching response. The agreement between numerical calculations and experimental measurements is remarkable.

This binary switching configuration with electromagnetically interacting high refractive index dimers opens the way to generate new and practical optical devices.

**Methods**

**Theoretical methods.** From the numerical point of view, the results are obtained by means of a Finite Element Method implemented in the commercial software COMSOL Multiphysics[36]. In particular, we use the Radio Frequency Module that allows us to formulate and solve the differential form of Maxwell's equations (in the frequency domain) together with boundary conditions. We thus take advantage of the far-field pattern computation module (see the Supplementary Materials for more details). The dimer is placed at the center of a spherical homogeneous region filled with air, whose radius is *λ/2+2R*. A perfectly matched layer (PML) domain, with thickness *λ/4*, is positioned outside of the embedding medium domain and acts as an absorber for the scattered field. The mesh is chosen sufficiently fine as to allow numerical convergence of the results. In particular, the element size of the mesh of the embedding medium is smaller than *λ/5* and that of the particles is smaller than *λ/(3\*Re(n))*.

In order to understand more deeply the underlying physics, we also theoretically separate the different terms that contribute to the scattered intensity thanks to a Green's function formalism. In that case, each component of the dimer is modeled by an electric dipole and a magnetic dipole which are at right angles to each other as well as perpendicular to the propagation direction of the incident wave. More details on the Green's function formalism and the dipolar approximation for the transverse configuration can be found in the Supplementary materials, while the derivation for the longitudinal configuration is provided in[37].

**Experimental methods**. Using the scale invariance rule, the so-called microwave analogy allows us to measure and characterize in the microwave range phenomena which are very promising in the optical domain, such as the sought-after "switching" effect. Thus the experiments described below are totally similar to those that could be observed (if possible) with dimers of silicon (Si) in the VIS-NIR range with nanosized dimensions of hundreds of nanometers. Our experiments were carried out using the microwave measurement facility in the anechoic chamber of the *Centre Commun de Ressources en Microondes* (CCRM) in Marseille, France. During the last decade, this facility has interestingly become a specialized microwave scattering device to perform analog to light measurements on a variety of complex particles[38]. It has recently been implemented in the scattering measurements on a single HRI subwavelength sphere to experimentally demonstrate the directional tunability of scattering radiations[19] and allowed the experimental proof of the Kerker conditions[13]. This experimental setup was used in this work to measure the scattered electric field, in both magnitude and phase and in parallel and perpendicular polarizations, by two analog-to-particle high refractive index spheres (Fig. 7).

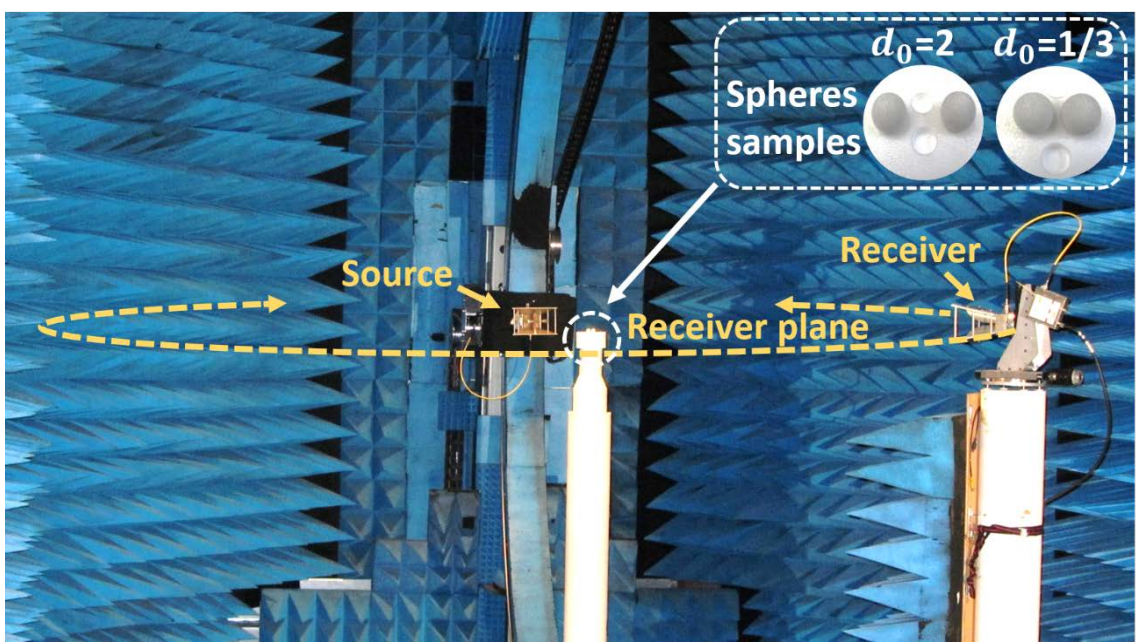

**Figure 7 | Experimental setup.** Picture of the experimental microwave setup in the anechoic chamber of the CCRM and the two HRI spheres in both large and small gap cases (in the inset). The source has a fixed position; the receiver moves circularly around the spheres which are positioned on the central vertical polystyrene mast. The scattered field measurement is performed in the horizontal scattering plane containing the source, the receiver and the target (spheres). The configuration in the photograph corresponds to the perpendicular polarization ($I_S$) case, here the transverse polarization configuration. The parallel polarization ($I_P$) case, here the longitudinal configuration, is obtained by rotating the source and the receiver.

The two HRI spheres are set at the center of a horizontal disk of around 4 m in diameter. They are placed on a vertical polystyrene mast, almost transparent to electromagnetic waves and illuminated by an incident plane wave emitted from a fixed-position source. Both spheres are made of the same low-loss high refractive index material (HIK 500F from Laird Technologies) with $R_1=R_2=R=9$ mm and $\varepsilon = 15.7 + 0.3i$. This permittivity was determined at the working frequencies (from 3 to 9 GHz) from far-field scattering measurements acquired with only one of the spheres and through comparison to Mie calculations, following the technique described in[39]. First, the scattered field at each receiving position is used to determine one permittivity value over all the frequencies, as the permittivity has been observed to have insignificant variations with respect to the frequency. Afterwards, out of the receiving positions where the measurements are of lowest errors and lowest sensitivity to noise, one averaged permittivity is estimated and is used in all the numerical simulations. In order to control the gaps between the spheres (3 mm and 18 mm representing the small and large gaps respectively), a polystyrene spheres-holder was especially fabricated for this purpose.

The experiments are not limited to the scattering measurement at 90° but are advantageously performed over an angular excursion of 260° around the target. This allows us to apply an angular-based post-processing procedure, important to reduce drift and noise effects and to enhance the data quality[40]. We draw the attention here to the challenges associated to this measurement, since the spheres have small sizes compared to the wavelength (small $q$), the scattered intensity is relatively low thus very sensitive to random noise, drift problems and parasite signals. Notice that all the measurements presented here (apart for Fig. 5) are calibrated which means that all the plotted intensities and values are truly quantitative ones (see the Supplementary Materials for more details).

**References**

1. Prasad, P.N. (ed.) Nanophotonics (John Wiley and Sons, 2004).

2. Kelly, K. L., Coronado, E., Zhao, L. L., & Schatz, G. The optical properties of metal nanoparticles: the influence of size, shape and dielectric environment. *J. Phys. Chem. B* **107**, 668–677 (2003).

3. Maier, S.A. Plasmonics: Fundamentals and Applications (Springer: Berlin Heidelberg, 2007).

4. Forouhi, A. & Bloomer, I. Optical properties of crystalline semiconductors and dielectrics. *Phys. Rev. B* **38**, 1865–1874 (1988).

5. García-Etxarri, A., Gómez-Medina, R., Froufe-Perez, L. S., López, C., Chantada, L., Scheffold, F., Aizpurua, J., Nieto-Vesperinas, M., & Sáenz, J. J. Strong magnetic response of submicron silicon particles in the infrared. *Opt. Express* **19**, 4815–4826 (2011).

6. Shi, L., Harris, J. T., Fenollosa, R., Rodriguez, I., Lu, X., Korgel, B. A., & Meseguer, F. Monodisperse silicon nanocavities and photonic crystals with magnetic response in the optical region. *Nat. Commun.* **4**, 1904 (2013).

7. Evlyukhin, A. B., Reinhardt, C., Seidel, A., Luk'yanchuk, B., & Chichkov, B. N. Optical response features of si-nanoparticle arrays. *Phys. Rev. B* **82**, 045404 (2010).

8. Gomez-Medina, R., García-Camara, B., Suárez-Lacalle, I., González, F., Moreno, F., Nieto-Vesperinas, M., & Sáenz, J. J. Electric and magnetic dipolar response of germanium nanospheres: interference effects, scattering anisotropy, and optical forces. *J. Nanophoton.* **5**, 053512 (2011).

9. Kuznetsov, A. I., Miroshnichenko, A. E., Fu, Y. H., Zhang, J., & Luk'yanchuk, B. (2012). Magnetic light. *Sci. Rep.* **2**, 492 (2012).

10. García-Cámara, B., Moreno, F., González, F., Saiz, J. M., & Videen, G. Light scattering resonances in small particles with electric and magnetic properties. *J. Opt. Soc. Am. A Opt. Image Sci. Vis.* **25**, 327–34 (2008).

11. Bakker, R. M., Permyakov, D., Yu, Y. F., Markovich D., Paniagua-Domínguez, R., Gonzaga, L., Samusev, A., Kivshar, Y., Luk'yanchuk, B., & Kuznetsov, A. I. Magnetic and Electric Hotspots with Silicon Nanodimers Magnetic and Electric Hotspots with Silicon Nanodimers. *Nano Lett.* **15**, 2137–2142 (2015).

12. Kerker, M., Wang, D., & Giles, C. L. Electromagnetic scattering by magnetic spheres. *J. Opt. Soc. Am.* **73**, 765–767 (1983).

13. Geffrin, J. M., García-Camara, B., Gómez-Medina, R., Albella, P., Froufe-Pérez, L., Eyraud, C., Litman, A., Vaillon, R., Gonzalez, F., Nieto-Vesperinas, M., Sáenz, J. J., & Moreno, F. Magnetic and electric coherence in forward- and back-scattered electromagnetic waves by a single dielectric subwavelength sphere," *Nat. Commun.* **3**, 1171 (2012).

14. Zambrana-Puyalto, X., Fernandez-Corbaton, I., Juan, M. L., Vidal, X., & Molina-Terriza, G. Duality symmetry and Kerker conditions. *Opt. Lett.* **38**, 1857–1859 (2013).

15. García-Cámara, B., Alcaraz de la Osa, R., Saiz, J. M., González, F., & Moreno, F. Directionality in scattering by nanoparticles: Kerker's null-scattering conditions revisited. *Opt. Lett.* **36**, 728–730 (2011).

16. Laee, R. A., Ilter, R. F., Ehr, D. L., & Ederer, F. L. A generalized Kerker condition for highly directive nanoantennas. *Opt. Lett.* **40**, 2645-2648 (2015).

17. Fu, Y. H., Kuznetsov, A. I., Miroshnichenko, A. E., Yu, Y. F. & Luk'yanchuk, B. Directional visible light scattering by silicon nanoparticles. *Nat. Commun.* **4**, 1527 (2013).

18. Zhang, Y., Nieto-Vesperinas, M., & Sáenz, J. J. Dielectric spheres with maximum forward scattering and zero backscattering: A search for their material composition. *J. Opt.* **17**, 105612-105615 (2013).

19. Tribelsky, M. I., Geffrin, J.-M., Litman, A., Eyraud, C., & Moreno, F. Small Dielectric Spheres with High Refractive Index as New Multifunctional Elements for Optical Devices. *Sci. Rep.* **5**, 12288 (2015).

20. García-Cámara, B., Algorri, J. F., Cuadrado, A., Urruchi, V., Sánchez-Pena, J. M., Serna, R. & Vergaz, R. All-Optical Nanometric Switch based on the Directional Scattering of Semiconductor Nanoparticles. *J. Phys. Chem. C.* **119**, 19558-19564 (2015).

21. García-Cámara, B., Algorri, J., Urruchi, V., & Sánchez-Pena, J. (2014). Directional Scattering of Semiconductor Nanoparticles Embedded in a Liquid Crystal. *Materials* **7**, 2784–2794 (2014).

22. Tribelsky, M. I. & Luk'yanchuk, B. Anomalous Light Scattering by Small Particles, *Phys. Rev. Lett.* **97**, 263902 (2006).

23. Yan, J., Liu, P., Lin, Z., Wang, H., Chen, H., Wang, C., & Yang, G. Directional Fano resonance in a silicon nanosphere dimer. *ACS Nano* **9**, 2968–2980 (2015).

24. Jia, Z. Y., Li, J. N., Wu, H. W., Wang, C., Chen, T. Y., Peng, R. W., & Wang, M. Dipole coupling and dual Fano resonances in a silicon nanodimer. *J. Appl. Phys.* **119**, 074302 (2016).

25. Staude, I., Miroshnichenko, A. E., Decker, M., Fofang, N. T., Liu, S., Gonzales, E., Dominguez, J., Luk, T. S., Neshev, D. N., Brener, I. & Kivshar, Y. Tailoring directional scattering through magnetic and electric resonances in subwavelength silicon nanodisks. *ACS Nano* **7**, 7824–7832 (2013).

26. Albella, P., Shibanuma, T., & Maier, S. A. Switchable directional scattering of electromagnetic radiation with subwavelength asymmetric silicon dimers. *Sci. Rep.* **5**, 18322 (2015).

27. Yan, J. H., Liu, P., Lin, Z. Y., Wang, H., Chen, H. J., Wang, C. X., & Yang, G. W. Magnetically induced forward scattering at visible wavelengths in silicon nanosphere oligomers. *Nat. Commun*. **6**, 7042 (2015).

28. García-Cámara, B., González, F., & Moreno, F. Linear polarization degree for detecting magnetic properties of small particles. *Opt. Lett*. **35**, 4084–4086 (2010).

29. García-Cámara, B., Gómez-Medina, R., González, F., Sáenz, J. J., Nieto-Vesperinas, M., & Moreno, F. Polarization analysis of the scattered radiation by silicon nanoparticles in the infrared. *AAPP* **89**, (S. 1) (2011).

30. Setién, B., Albella, P., Saiz, J. M., González, F. & Moreno, F. *New J. Phys*. **12**, 103031 (2010).

31. Bohren, C.F. & Huffman, D.R. Absorption and Scattering of Light by Small Particles (John Wiley and Sons, 1983).

32. García-Camara, B., Gómez-Medina, R., Sáenz, J. J. & Sepúlveda, B. Sensing with magnetic dipolar resonances in semiconductor nanospheres. *Opt. Express* **21**, 23007–23020 (2013).

33. Barreda, A. I., Sanz, J. M., & González, F. Using linear polarization for sensing and sizing dielectric nanoparticles. *Opt. Express* **23**, 9157-9166 (2015).

34. Barreda, A. I., Sanz, J. M., Alcaraz de la Osa, R., Saiz, J. M., Moreno, F., González, F. & Videen, G. *J. Quant. Spectrosc. Radiat. Transfer* **162**, 190-196 (2015).

35. Meier, M., & Wokaun, A. Enhanced fields on large metal particles: dynamic depolarization. *Opt. Lett.* **8**, 581–583 (1983).

36. COMSOL Multiphysics 5.0; Comsol Inc.: Burlington, MA, 2015.

37. Albella, P., Poyli, M. A., Schmidt, M. K., Maier, S. A., Moreno, F., Sáenz, J. J., & Aizpurua, J. Low-Loss Electric and Magnetic Field-Enhanced Spectroscopy with Subwavelength Silicon Dimers. *J. Phys. Chem. C* **117**, 13573−13584 (2013).

38. Vaillon, R., & Geffrin, J.M. Recent advances in microwave analog to light scattering experiments. Journal of Quantitative Spectroscopy and Radiative Transfer, **146**, 100-105 (2014) 39. Eyraud, C., Geffrin, J.M., Litman, A., & Tortel, H. Complex Permittivity Determination from Far-Field Scattering Patterns. IEEE Antennas and Wireless Propagation Letters, **14**, 309-312 (2015).

40. Eyraud, C., Geffrin, J.M., Litman, A., Sabouroux, P., & Giovannini, H. Drift correction for scattering measurements, **89,** 244104 (2006).

**Acknowledgements**

This research has been supported by MICINN (Spanish Ministry of Science and Innovation, project FIS2013-45854-P) and Fundación IBERDROLA España (Research on Energy and the Environment Grants). AIB wants to thank the University of Cantabria for her FPU grant. We also acknowledge the opportunity provided by the *Centre Commun de Ressources en Microonde* to use its fully equipped anechoic chamber and for financing HS PhD grant.

**Author contributions**

JMG, AB, HS and FM conceived this work. AB, FG, AL and FM developed the concept. JMG, FG, HS and FM conceived the experimental realization in the microwave regime. AB, AL and FM performed the theoretical background. JMG and HS designed the experimental setup and performed the experiments. JMG, AL and HS developed the post-processing treatments of the experimental data. HS, AB, JMG, AL, FG and FM analyzed the experimental data. AB, AL, HS and JMG carried out numerical calculations and figures. AB, AL, JMG and FM wrote the paper. All authors contributed to scientific discussion and critical revision of the article. FM and AL supervised the study.

**Competing financial interests**. The authors declare no competing financial interests.

**Materials & Correspondence**. Fernando Moreno. morenof@unican.es